\documentclass{article}
\usepackage{frascatiphys_C}
\usepackage{epsfig}
\begin{document}
\title{ 
FUTURE KAON PROGRAM AT KEK/J-PARC
}
\author{
 Takeshi K. Komatsubara\\
\em High Energy Accelerator Research Organization (KEK),
    Ibaraki 305-0801, Japan\\
Email: takeshi.komatsubara@kek.jp
}
\maketitle
\baselineskip=11.6pt
\begin{abstract}
The current program of kaon-decay experiments 
at the KEK 12~GeV Proton Synchrotron (KEK-PS) and 
the prospects for the future kaon program at the
new 50~GeV PS of J-PARC, being constructed in Japan, 
are reviewed. 
\end{abstract}
\baselineskip=14pt
\section{Overview}
 Experiments at KEK-PS started in 1977, and  
distinguished kaon experiments in search of 
$K^+\to\pi^+\nu\bar{\nu}$, 
heavy-neutrino emission in $K^+\to\mu^+\nu$, 
right-handed currents in $K^+\to\mu^+\nu$
and $K^0_L\to\mu^{\pm} e^{\mp}$, 
respectively, were made in 1980's. 
After the Booster of BNL-AGS increased the 
proton intensity to be high, 
a measurement of $K^0_L\to\pi^+\pi^- e^+ e^-$
and a search for T-violating transverse muon polarization 
in $K^+\to\pi^0\mu^+\nu$, 
which were suitable for low-energy kaons 
and complementary to the experiments in other laboratories,
were performed at KEK-PS.
Kaon physicists also participated 
in the E787/E949 experiments at BNL 
and the KTeV experiment at FNAL through the Japan-U.S. 
Cooperative Research Program. 

 To this day KEK-PS delivers fast-extracted beams to
 the K2K long-baseline neutrino experiment for 6 months per year
 and slow-extracted beams to the experiments 
 in the East and North Counter Halls for 2 to 4 months per year~\cite{KEKPS}. 
 A typical slow-extracted beam is $2.5\cdot 10^{12}$ protons per 
 2.0-second spill in every 4.0 seconds. 
 Experiment 391a~\cite{E391a}, which is the first dedicated search 
 for the $K^0_L\to\pi^0\nu\bar{\nu}$ decay, 
 has carried out the first physics run  successfully 
 from February to June 2004. 

\begin{figure}[t]
 \vspace{2.0cm}
 \epsfig{file=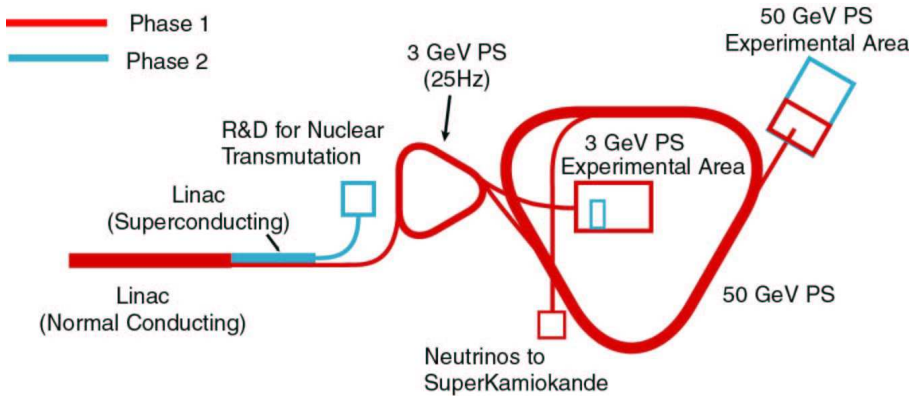,width=1.00\linewidth}
 \caption{\it
      J-PARC accelerators.
    \label{JPARC} }
\end{figure}  
  J-PARC, which stands for 
  Japan Proton Accelerator Research Complex~\cite{J-PARC}, 
  is the joint project of 
  Japan Atomic Energy Research Institute (JAERI) and KEK. 
  The accelerators (fig.\ref{JPARC}) are under construction
  at the Tokai site of JAERI located at 50km northeast of KEK.
  The construction will be finished in 2008 and, 
  with very intense proton beams from the new 50~GeV PS, 
  great opportunities for various researches in nuclear and particle 
  physics, including kaon experiments with much higher sensitivities 
  than ever, would be opened. 

  The rest of this article is devoted to a report of
  the status of the E391a experiment 
  and the future kaon experiments at J-PARC.  
  The E246/E470 experiments on T-violation in $K^+\to\pi^0\mu^+\nu$
  and direct photon emission in $K^+\to\pi^+\pi^0\gamma$ are 
  reported elsewhere~\cite{E246}~\cite{E470}. 

\section{E391a Experiment for $K^0_L\to\pi^0\nu\bar{\nu}$}
 Observation of the rare decay $K^0_L\to\pi^0\nu\bar{\nu}$ is 
 a new evidence for CP violation in kaon decays. 
 The branching ratio is represented within the Standard Model (SM) 
 as~\cite{Buras-review}:
\begin{equation}
 B(K^0_L\to\pi^0\nu\bar{\nu}) = \\
  2.12 \cdot 10^{-10} \times [\ \frac{\lambda}{0.224}\ ]^8 \times 
      (\ \frac{Im\lambda_t}{\lambda^5} \cdot X(x_{t})\ )^{2}
\end{equation}
 where $X(x_{t})$ is the Inami-Lim loop function~\cite{InamiLim}
 with the QCD correction, 
  $x_t$ is the square of the ratio of the top to W masses, and
\begin{equation}
  \lambda_{t} \equiv V_{ts}^{*} \cdot V_{td} =  
      A^{2} \lambda^5 \cdot (1 - \rho - i \eta)
\end{equation}
in the Wolfenstein parametrization $A$, $\lambda$, $\rho$, and $\eta$.
The SM prediction is 
$(3.0\pm 0.6)\cdot 10^{-11}$, in which theoretical uncertainties
are only a few \%.
A model-independent bound called the Grossman-Nir limit~\cite{GrossmanNir}:
\begin{equation}
 B(K^0_L\to\pi^0\nu\bar{\nu}) < 
  4.4 \times B(K^+\to\pi^+\nu\bar{\nu})
   < 1.4 \cdot 10^{-9}
\end{equation}
can be extracted 
from its isospin-relation to 
the $K^+\to\pi^+\nu\bar{\nu}$ decay~\cite{E949}.
New physics beyond the SM 
could enhance the branching ratio by one order of magnitude:  
$(3.1\pm 1.0)\cdot 10^{-10}$~\cite{Buras-new}.
The current upper limit on the branching ratio 
$< 5.9 \cdot 10^{-7}$ was set 
by the KTEV collaboration\cite{KTEVpinndalitz} 
using the Dalitz decay mode $\pi^0\to e^+e^-\gamma$ of 1.2\%.

\begin{figure}[ht]
 \vspace{-1.0cm}
 \epsfig{file=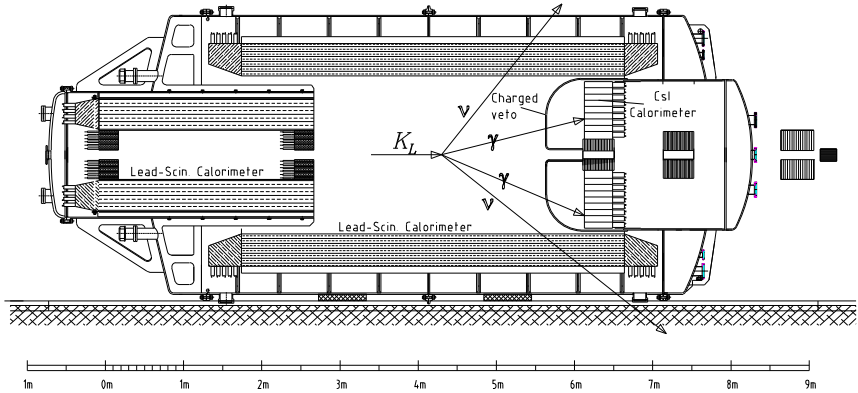,width=0.95\linewidth}
  \caption{\it
    Side-view of the E391a detector.
    \label{e391adet} }
\end{figure}
The E391a experiment~\footnote{
 E391a is an international collaboration of 
 KEK, Saga, Yamagata, RCNP, Osaka, NDA, 
 JINR, Chicago, TNU, Pusan, and Kyoto.}(fig.\ref{e391adet})
searches for the $K^0_L\to\pi^0\nu\bar{\nu}$ decay
with collimated ``pencil'' neutral beams.
An endcap calorimeter 
with undoped CsI crystals detects two photons from $\pi^0\to\gamma\gamma$
and measures their energy and position.
The $K^0_L$-decay vertex position along the beam line 
is determined from the constraint of $\pi^0$ mass. 
Calorimeters that cover the decay region do
hermetic photon detection and reject the background from $K^0_L\to\pi^0\pi^0$.
Charged particles are removed 
by their energy deposits in a plastic scintillator
in front of each calorimeter.

\begin{figure}[t]
 \epsfig{file=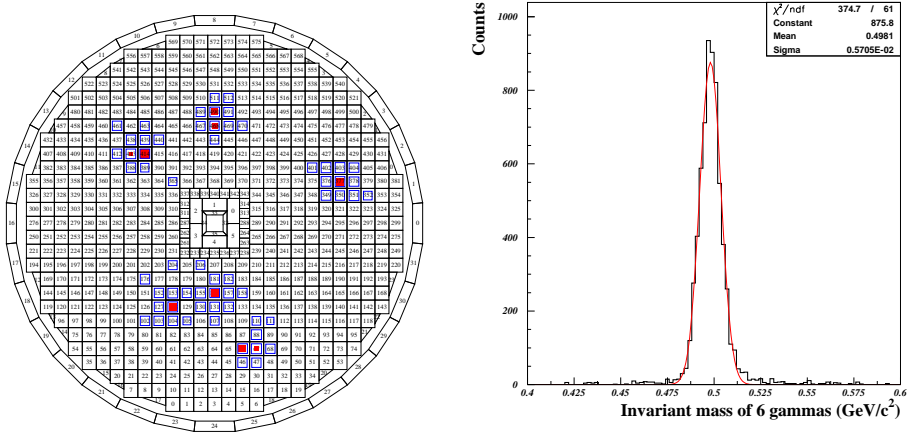,width=1.00\linewidth}
 \caption{\it
      Display of an event with six clusters 
      in the E391a CsI calorimeter (left); 
      invariant-mass distribution of the $K^0_L\to\pi^0 \pi^0 \pi^0$ decay 
      reconstructed from the six-cluster events
      (right). 
    \label{e391a-6cls} }
\end{figure}  
Beam line survey and detector construction were performed 
from 2001 to 2003, 
and the first physics run was carried out in 2004. 
The beam line, which had been designed carefully, 
 provided clean neutral beams; 
 in the decay region  
 a high vacuum of $1.21\cdot 10^{-5}$ Pa was achieved. 
Fig.\ref{e391a-6cls} shows the $K^0_L\to\pi^0 \pi^0 \pi^0$ decay reconstructed 
from the events with six clusters in the CsI calorimeter. 
These events were used online to monitor the beam line and 
detector during the data taking. 

The goal of E391a is to achieve a sensitivity 
below the Grossman-Nir limit ($1.4 \cdot 10^{-9}$)
and to reach the level predicted by new-physics 
($3.1\cdot 10^{-10}$).
The analysis is in progress. 
They plan to continue the study at J-PARC.

\section{Future Kaon Experiments at J-PARC}

 The J-PARC 50~GeV PS was designed to provide,
 in the slow extraction, 
 $300\cdot 10^{12}$ protons per 0.7-second spill 
 in every 3.42 seconds to an experimental area 
 named Hadron Experimental Hall.
 The beam energy at the initial operation phase (Phase-1) will be 
 30~GeV. 

 Call for Letters of Intent (LoI's) for nuclear and particle physics 
 experiments at the J-PARC was issued in July 2002, and 
 thirty LoI's~\cite{LoIs} were submitted. 
 There were five LoI's for kaon experiments: 
     \begin{itemize}
        \item measurement of the $K^0_L\to\pi^0\nu\bar{\nu}$ branching ratio
     \end{itemize}
 with neutral beams and 
      \begin{itemize}
         \item study of the $K^+\to\pi^+\nu\bar{\nu}$ decay,
         \item search for T-violation in $K^+$ decay,
         \item study of the decay spectra of stopped kaons, and
         \item precise measurement of the 
               $K^+\to\pi^0 e^+ \nu$ branching ratio
      \end{itemize}
 with $K^+$ beams of low momentum (0.6-0.8~GeV/$c$).
 These LoI's are regarded as a natural extension of 
 the kaon program that has been worked out 
 (E391a, BNL-E949 and E246/E470).
 In the beam-line layout plan
  of the Hadron Experimental Hall at Phase-1, 
  reported in February 2004~\cite{Hall},  
 the hall has been designed so as to accommodate these experiments
 in the future. 
 Call for full proposals is expected to be issued in the autumn of 2004; 
 intensive discussions have been held in a series of workshops~\cite{NP}.

\section*{Acknowledgments}
I would like to thank J.~Imazato, T.~Inagaki, G.Y.~Lim, S.~Sugimoto
and T.~Yamanaka for useful discussions.
I would like to acknowledge support 
from Grant-in-Aid for Scientific Research 
in Priority Area: ``Mass Origin and Supersymmetry Physics''
by the MEXT Ministry of Japan.
\end{document}